\documentclass[11pt]{article}

\advance \textheight by \topskip
\makeatletter
\@addtoreset{equation}{section}
    
\makeatother

\usepackage{amsmath,amssymb}

\begin{document}

\title{
{\bf   Electric Chern-Simons term, enlarged exotic Galilei symmetry
and noncommutative plane}}

\author
{{\sf Mariano  A. del Olmo${}^a$}\thanks{E-mail: olmo@fta.uva.es}
{\sf\ and Mikhail S. Plyushchay${}^b$}\thanks{E-mail:
mplyushc@lauca.usach.cl}
\\[4pt]
{\small \it ${}^a$Departamento de F\'{\i}sica,
Universidad de Valladolid}\\
{\small \it E-47011, Valladolid, Spain}\\
 {\small \it ${}^b$Departamento de F\'{\i}sica,
Universidad de Santiago de Chile}\\
{\small \it Casilla 307, Santiago 2, Chile}\\
}
\date{}

\maketitle

\begin{abstract}
The extended exotic planar model  for a charged particle is
constructed. It includes a Chern-Simons-like term for a dynamical
electric field, but produces usual equations of motion for the
particle in background constant uniform electric and magnetic
fields. The electric Chern-Simons term is responsible for the
non-commutativity of the boost generators in the ten-dimensional
enlarged exotic Galilei symmetry algebra of the extended system. The
model admits two reduction schemes by the integrals of motion, one
of which reproduces the usual formulation for the charged particle
in external constant electric and magnetic fields with associated
field-deformed Galilei symmetry, whose commuting boost generators
are identified with the nonlocal in time Noether charges reduced
on-shell. Another reduction scheme, in which electric field
transmutes into the commuting space translation generators, extracts
from the model a free particle on the noncommutative plane described
by the two-fold centrally extended Galilei group of the
non-relativistic anyons.
\end{abstract}

\vskip.5cm\noindent

\section{Introduction}

Nowadays, the questions on  symmetries of noncommutative  systems
and their relation to the  Poincar\'e and Galilei symmetries are the
hot topics \cite{NCS1,NCS2,NCS3,NCS4,Luk,Az,LSZ,DH,JN,HP1}. A simple
example of such a system is a model of a particle in a
noncommutative plane  \cite{LSZ,DH,JN,HP1,HP2,HP3}. In a free case
the model is described by the two-fold centrally extended exotic
Galilei symmetry with noncommuting boost generators \cite{LL}.
Recently it was observed that a \emph{free} non-relativistic anyon
(NR-anyon) can be identified with a free planar particle with
noncommuting coordinates \cite{JN,HP1,HP2}. In its evolution the
NR-anyon reveals, like a Dirac particle, a nonrelativistic
Zitterbewegung superimposed on a translation motion \cite{HP1,HP2}.
As a result, the free NR-anyon motion turns out to be similar to the
evolution law of a charged particle in external constant homogeneous
electric and magnetic fields (further on, EM-particle), with a
translation motion of the NR-anyon recognized as an analog of a Hall
drift motion of the EM-particle. It is well known that the
EM-particle model also reveals the elements of noncommutative
geometry: its guiding center coordinates are noncommuting. However,
unlike the free NR-anyon, the EM-particle is described  by the
broken, field-deformed Galilei symmetry with noncommuting
translation generators \cite{AzIz}. Therefore, a natural question
appears on the relation between these two essentially different
systems and their symmetries in the light of the dynamical
similarity.

In the present paper, we construct the extended exotic model
(\emph{Ex}-model), which unifies the both systems in a nontrivial
way. The ``unification'' is achieved by making the electric field to
be dynamical, and by including
an electric  Chern-Simons-like term  into Lagrangian of the EM-particle
(see Eq. (\ref{LCen})).

The obtained system  is characterized by the following properties.
The extension of the model does not change equations of motion of
the initial EM-particle but transforms its nonlocal Noether charges
associated with the Galilei boost and rotation symmetries into the
local integrals of motion. Lagrangian (\ref{LCen}) of the extended
system looks formally as a sum of Lagrangians of the EM-particle and
of the NR-anyon of \emph{zero} mass, with electric field playing the
role of the momentum  for the latter subsystem. The two subsystems
are coupled, however, by means of the electric interaction term of
the first subsystem and of the kinetic term for the  second
subsystem coordinates conjugate to the electric field variables.
These two terms  form together the Lagrangian constraint, which
identifies the \emph{commutative} configuration ($x$-) space of the
first subsystem to be tangent to the \emph{noncommutative}
configuration ($q$-) space of the second subsystem. The switching
off interaction in the first subsystem by putting the particle's
charge to be equal to zero, reduces the extended system to the
decoupled sum of the two free non-relativistic planar models
--- the usual massive scalar particle and the free exotic particle of Duval and
Horvathy \cite{DH} of zero mass.

The symmetry of the extended model can be  identified as a
ten-dimensional enlarged exotic Galilei group. In comparison with
the planar Galilei-Maxwell symmetry discussed earlier \cite{NO,NOT},
it includes two additional generators associated with the circular
motion of the EM-particle, whose analogs in the NR-anyon model are
responsible for the Zitterbewegung. The electric Chern-Simons term
induces the non-commutativity of the Galilei boost generators of the
\emph{Ex}-model. The Galilei symmetry acts in the  $x$- as well as
in $q$- spaces, while the electric field, responsible for the Hall
drift motion in the $x$-space, acts nontrivially only in the
$q$-space, where it generates translations. Additional circular
motion vector integral is invariant with respect to the boosts and
$x$- and $q$- translations.

The obtained system admits two reduction schemes by the integrals of
motion, one of which reproduces the EM-particle with associated
field-deformed Galilei symmetry, whose commuting boost generators
are identified with the nonlocal in time Noether charges reduced
on-shell. Another reduction scheme, in which the electric field
transmutes into the commuting space translation generators, extracts
from the model a free massive particle on the noncommutative plane
\cite{DH} described by the two-fold centrally extended Galilei group
of the non-relativistic anyons.

The paper is organized as follows. In Section 2 we analyze the
minimal and extended formulations for the free non-relativistic
anyon, and observe the enlargement of the exotic planar Galilei
symmetry by the integrals of motion associated with the
Zitterbewegung. In the third Section we consider the charged
particle in external constant uniform electric and magnetic fields.
We compare the dynamics of the model with that of the free NR-anyon
in extended formulation, and analyze in detail its symmetries
corresponding to the field-deformed Galilei algebra. In particular,
we observe how the \emph{nonlocal} Noether charges associated with
the boosts and rotations are transformed on shell into the
\emph{local} integrals of motion generating the complicated
field-deformed symmetry transformations, which take the form of the
usual Galilei boosts and rotations under switching off the
electromagnetic coupling. In the fourth Section, we construct the
extended exotic model and analyze its symmetries.  In Section 5 we
consider the two Hamiltonian reduction schemes, which relate the
extended system to the EM-particle and to the NR-anyon models. The
last Section is devoted to the discussion and  concluding remarks.

\section{NR-anyon: symmetries, motion,  coordinates}
The exotic two-fold centrally extended planar Galilei group
\cite{LL,DH,HP2} can be given by the Poisson bracket relations of the
generators of the space, ${\cal P}_i$, and time, ${\cal H}$,
translations, of the rotations, ${\cal J}$, and boosts, ${\cal
K}_i$,
\begin{eqnarray}
&\{{\cal P}_i,{\cal P}_j\}=0,\qquad \{{\cal K}_i,{\cal
P}_j\}=m\delta_{ij},\qquad \{{\cal K}_i,{\cal
K}_j\}=-\kappa\epsilon_{ij},&\nonumber
\\
&\{{\cal J},{\cal P}_i\}=\epsilon_{ij}{\cal P}_j,\qquad \{{\cal
J},{\cal K}_i\}=\epsilon_{ij}{\cal K}_j,&\label{PPK}
\\
&\{{\cal H},{\cal K}_i\}=-{\cal P}_i,\qquad \{{\cal H},{\cal
P}_i\}=0,\qquad \{{\cal H},{\cal J}\}=0.&\nonumber
\end{eqnarray}
The parameters $m$ and $\kappa$ play the role of the two central
charges of the algebra (\ref{PPK}), characterized by the two
Casimir elements \cite{HP2}
\begin{equation}
{\cal C}_1=\epsilon_{ij}{\cal K}_i{\cal P}_j-m{\cal J}-
\kappa{\cal H},\qquad
{\cal C}_2={\cal P}_i^2-2m{\cal H}.
\label{c1c2}
\end{equation}

There are two different but related possibilities to realize the
algebra (\ref{PPK}) as a symmetry of the NR-anyon: the minimal and
the extended ones\footnote{ Cf. the two formulations for (2+1)D
relativistic anyons \cite{RA,JN1,CP}.}. In view of the first
relation from (\ref{PPK}), the space translation generators can be
identified as particle's canonical momenta, ${\cal P}_i=p_i$.
Then, in minimal realization the coordinates of the particle are
reconstructed from the boost and space translation generators,
\begin{equation}
{\cal X}_i= \frac{1}{m}{\cal K}_i-\theta\epsilon_{ij}{\cal P}_j
+\frac{1}{m}{\cal P}_i\, t,
 \label{calX}
\end{equation}
where
$$
\theta=\frac{\kappa}{m^2}.
$$
 This guarantees the Galilei-covariant properties of the
${\cal X}_i$, as well as its free evolution law:
$$ \{{\cal J},{\cal
X}_i\}=\epsilon_{ij}{\cal X}_j,\qquad \{{\cal P}_i,{\cal
X}_j\}=-\delta_{ij},\qquad \{{\cal K}_i,{\cal X}_j\}=t\delta_{ij},
$$
$$
\frac{d}{dt}{\cal X}_i=\frac{\partial}{\partial t}{\cal X}_i
+\{{\cal X}_i,{\cal H}\}=\frac{1}{m}{\cal P}_i.
$$
Fixing the values of the Casimir elements (\ref{c1c2}) to be equal
to zero, we get the free massive planar particle with zero internal energy
and spin. Its symmetry generators, represented in terms of the
variables ${\cal X}_i$ and $p_i$, are
\begin{equation}\label{KX}
{\cal P}_i=p_i,\qquad {\cal K}_i=m{\cal
X}_i-tp_i+m\theta\epsilon_{ij}p_j,\qquad {\cal
J}=\epsilon_{ij}{\cal X}_ip_j+\frac{1}{2}\theta p_i^2, \qquad
{\cal H}=\frac{1}{2m}p_i^2.
\end{equation}
The price we pay for such a minimal realization is the
noncommutativity of the coordinates ${\cal X}_i$. In
correspondence with Eqs. (\ref{calX}), (\ref{PPK}), the
fundamental phase space Poisson bracket relations are
\begin{equation}\label{calXp}
\{{\cal X}_i,{\cal X}_j\}=\theta\epsilon_{ij},
\qquad
\{{\cal X}_i,p_j\}=\delta_{ij},\qquad
\{p_i,p_j\}=0.
\end{equation}
One can also define another set of noncommuting coordinates
\cite{HP2,HP3},
\begin{equation}\label{Y}
{\cal Y}_i = {\cal X}_i+\theta \epsilon_{ij}p_j, \qquad \{{\cal
Y}_i,{\cal Y}_j\}=-\theta\epsilon_{ij}, \qquad \{{\cal Y}_i,{\cal
X}_j\}=0,\qquad \{{\cal Y}_i,p_j\}=\delta_{ij},
\end{equation}
which simplify the form of the boost and rotation generators,
\begin{equation}\label{KJY}
{\cal K}_i=m{\cal Y}_i-tp_i,\qquad {\cal J}=\frac{1}{2\theta}
\left({\cal Y}_i^2-{\cal X}_i^2\right).
\end{equation}
Since
\begin{equation}\label{KPJY}
\{{\cal J},{\cal Y}_i\}=\epsilon_{ij}{\cal Y}_j,\qquad
 \{{\cal
P}_i,{\cal Y}_j\}=-\delta_{ij},\qquad \{{\cal K}_i,{\cal
Y}_j\}=t\delta_{ij}-m\theta\epsilon_{ij},
\end{equation}
the last relation means that the coordinates ${\cal Y}_i$, unlike
the ${\cal X}_i$, are \emph{not} Galilei-covariant.

At the quantum level, the ${\cal X}_i$, $i=1,2$,  are the
non-commutative (hence, simultaneously not localizable)
coordinates. One can work there in the momentum representation.
Or, by introducing commuting (canonical) coordinates
\begin{equation}\label{mX}
{\mathfrak X}_i= \frac{1}{2}({\cal X}_i+{\cal Y}_i)={\cal
X}_i-\frac{1}{2}\theta\epsilon_{ij}p_j,
\end{equation}
$\{{\mathfrak X}_i,{\mathfrak X}_j\}=0$, $\{{\mathfrak
X}_i,p_j\}=\delta_{ij}$, we get the coordinate representation
$\psi({\mathfrak X})$. In the latter case,
in correspondence with (\ref{mX}), the covariant coordinate
operators are realized by the star-product \cite{star},
$$
\hat{\cal X}_j\psi({\mathfrak X})=\left({\mathfrak
X}_j-\frac{i}{2}\theta\epsilon_{ jk}
{\partial}_k\right)\psi({\mathfrak X})={\mathfrak X}_j\star
\psi({\mathfrak X}) .
$$
In terms of the coordinates ${\mathfrak X}_i$, the angular
momentum takes a usual form, ${\cal J}=\epsilon_{ij}{\mathfrak
X}_ip_j$. However,  the canonical coordinates ${\mathfrak X}_i$,
like the ${\cal Y}_i$, are not Galilei-covariant, $\{{\cal K}_i,
{\mathfrak X}_j\}=t\delta_{ij}-\frac{1}{2}m\theta\epsilon_{ij}$.

The covariance and localizability may simultaneously be
incorporated into the theory by introducing the Galilei-covariant
canonical coordinates $x_i$ alongside with the
translation-invariant internal noncommuting variables $v_i$ being
analogs of the gamma matrices for the Dirac particle,
$$
\{x_i,x_j\}=0,\qquad
\{x_i,p_j\}=\delta_{ij},\qquad
\{v_i,v_j\}=-\kappa^{-1}\epsilon_{ij},\qquad
\{v_i,x_j\}=\{v_i,p_j\}=0.
$$
Taking into account relations (\ref{PPK}), the boost and rotation
generators are realized  in such an extended approach in the form
\cite{HP2}
\begin{equation}\label{kjkappa}
{\cal K}_i=mx_i-tp_i+\kappa\epsilon_{ij}v_j,\qquad
{\cal J}=\epsilon_{ij}x_ip_j+\frac{1}{2}\kappa
v_i^2.
\end{equation}
Then, fixing zero value for the first Casimir central element
(\ref{c1c2}), ${\cal C}_1=0$, we obtain the Hamiltonian,
\begin{equation}
{\cal H}=p_iv_i-\frac{1}{2}v_i^2. \label{hvv}
\end{equation}
It generates the equations of motion
\begin{equation}
\frac{dx_i}{dt}=v_i,\qquad \frac{dp_i}{dt}=0,\qquad
\frac{dv_i}{dt}=\omega\epsilon_{ij}(v_j-m^{-1}p_j), \label{xvp}
\end{equation}
where $\omega=\frac{m}{\kappa}$.
 Their solution is given by
\begin{equation}\label{solnc}
p_i(t)=\bar{p}_i,\qquad
x_i(t)=\frac{1}{m}\left(p_it-\epsilon_{ij}(V_j(t)- \bar{V}_j)
\right) +\bar{x}_i,\qquad
 V_i(t)=\bar{V}_j\Delta_{ji}(t),
\end{equation}
with
\begin{equation}
V_i= \kappa(v_i-m^{-1}p_i), \label{Vvp}
\end{equation}
$\bar{x}_i=x_i(0)$, $\bar{p}_i=p_i(0)$, $\bar{V}_i=V_i(0)$. Here
we have introduced the notation
\begin{equation}
\Delta_{ij}(t)=\delta_{ij}\cos\omega t- \epsilon_{ij}\sin\omega t=
\Delta_{ji}(-t)=\Delta_{ij}^{-1}(t) \label{Del}
\end{equation}
for the elements of the orthogonal matrix satisfying the equation
$\frac{d}{dt}\Delta_{ij}=-\omega\epsilon_{ik}\Delta_{kj}$.

 As in the case of the Dirac particle, the commuting
Galilei-covariant coordinates $x_i$ are subjected to the
Zitterbewegung: here the evolution is described by a superposition
of the translation and rotation motions. The $V_i$ are invariant
under the space translations and Galilei boosts, $\{p_i,V_j\}=0$,
$\{{\cal K}_i,V_j\}=0$. Therefore, the
\begin{equation}
{\cal X}_i=x_i+\frac{1}{m}\epsilon_{ij}V_j \label{XxV}
\end{equation}
is a Zitterbewegung-free Galilei-covariant vector satisfying the
equation of motion  $\frac{d}{dt}{\cal X}_i=\frac{1}{m}p_i$. It
has noncommuting components, $\{{\cal X}_i,{\cal
X}_j\}=\theta\epsilon_{ij}$, and can be identified with the
covariant coordinates (\ref{calX}) in the minimal formulation of
the free NR-anyon. This is an analog of the Foldy-Wouthuysen
coordinate for the Dirac particle. The analog of the
Galilei-noncovariant vector (\ref{Y}) is given here by ${\cal
Y}_i=x_i+m\theta\epsilon_{ij}v_j$.

  In accordance with
Eqs. (\ref{Vvp}), (\ref{XxV}), the transition from the phase space
variables $x_i$, $p_i$, $v_i$ to the set ${\cal X}_i$, $p_i$ and
$V_i$,

$$
\{{\cal X}_i,{\cal X}_j\}=\theta\epsilon_{ij},\qquad \{{\cal
X}_i,p_j\}=\delta_{ij},\qquad
\{V_i,V_j\}=-\kappa\epsilon_{ij},\qquad \{V_i,{\cal
X}_j\}=\{V_i,p_j\}=0,
$$
gives the following realization for the generators of the exotic
Galilei group:
\begin{equation}
{\cal P}_i=p_i,\qquad {\cal K}_i=m{\cal X}_i-tp_i+m\theta
\epsilon_{ij}p_j,\qquad {\cal J}=\epsilon_{ij}{\cal X}_ip_j
+\frac{1}{2}\theta p_i^2+\frac{1}{2\kappa} V_i^2, \label{kjvv}
\end{equation}
\begin{equation}
{\cal
H}=\frac{1}{2m}\left(p_i^2-\omega^{2}V_i^2\right).\label{hvV}
\end{equation}
The obtained system realizes the exotic planar Galilei group as
its symmetry, but it contains additional internal degrees of
freedom described by the $V_i$. The set of the Galilei group
generators can be extended here by the vector integral
\begin{equation}\label{VV}
{\cal V}_i=\Delta_{ij}(t)V_j= \bar{V}_i,
\end{equation}
for which  we have the relations
\begin{equation}\label{GalV}
 \{{\cal
V}_i,{\cal V}_j\}=-\kappa\epsilon_{ij},\qquad
 \{{\cal
H},{\cal V}_i\}=-\omega\epsilon_{ij}{\cal V}_j,\qquad \{{\cal
J},{\cal V}_i\}=\epsilon_{ij}{\cal V}_j, \qquad
\end{equation}
\begin{equation}\label{JVV}
\{{\cal P}_i,{\cal V}_j\}=\{{\cal K}_i,{\cal V}_j\}=0.
\end{equation}
The extended Galilei algebra with the Poisson bracket relations
(\ref{PPK}), (\ref{GalV}), (\ref{JVV}) is characterized by the
Casimir elements
\begin{equation}
{\cal C}_1=\epsilon_{ij}{\cal K}_i{\cal
P}_j+\frac{1}{2\kappa}{\cal V}_i^2-m{\cal J}- \kappa{\cal
H},\qquad {\cal C}_2={\cal P}_i^2-\omega^2{\cal V}_i^2-2m{\cal H}.
\label{C1C2}
\end{equation}
The comparison of Eq. (\ref{C1C2}) with Eq. (\ref{c1c2}) shows
that the free exotic particle in the minimal  formulation is
nothing else as the extended model reduced to  the surface ${\cal
V}_i=0$, $i=1,2$, which can equivalently be given by the two
second class phase space constraints not depending explicitly on
time,
\begin{equation}\label{VV2}
V_i\approx 0.
\end{equation}
The momenta $p_i$ as well as the covariant coordinates (\ref{XxV})
have zero Poisson brackets with constraints (\ref{VV2}).
After reduction coordinates (\ref{XxV}) are transformed into the
noncommuting coordinates of the minimal formulation with Dirac
bracket relations of the form (\ref{calXp}).

In correspondence with Dirac theory of the constrained systems, if
$\varphi_a\approx 0$ is a set of second class constraints, one can
define the extension $A^*$ of a dynamical quantity $A$,
\begin{equation}\label{A*}
A^*=A-\{A,\varphi_a\}C^{-1}_{ab}\varphi_b,\quad
C_{ab}=\{\varphi_a,\varphi_b\}.
\end{equation}
It satisfies the properties $\varphi_a^*=0$,
$\{A^*,\varphi_a\}\approx 0$, $\{A^*,D^*\}\approx \{A,D\}^*$, where
$\{A,D\}^*=\{A,D\}-\{A,\varphi_a\}C^{-1}_{ab}\{\varphi_b,D\}$ is the
Dirac bracket \cite{Dirac}. In this sense, we have $V_i^*=0$,
$p_i^*=p_i$ and $x_i^*={\cal X}_i$, where ${\cal X}_i$ is  the
Zitterbewegung-free covariant coordinate (\ref{XxV}).

Let us summarize shortly the NR-anyon model in the symplectic geometry
language \cite{FJ,Dirac}. The symplectic  two-form of the model in
the extended formulation can be presented in one of the equivalent
forms,
\begin{eqnarray}
\sigma&=&dp_i\wedge dx_i +\frac{1}{2}\epsilon_{ij}dv_i\wedge
dv_j\nonumber\\
&=&dp_i\wedge d{\cal X}_i+\frac{1}{2}\theta dp_i\wedge dp_j +
\frac{1}{\kappa}\epsilon_{ij}dV_i\wedge dV_j\nonumber\\
&=&\frac{1}{2\theta}\epsilon_{ij}(d{\cal Y}_i\wedge d{\cal
Y}_j-d{\cal X}\wedge d {\cal X}_j)+\frac{1}{2\kappa}
\epsilon_{ij}dV_i\wedge dV_j. \label{sigmaex}
\end{eqnarray}
The equations of motion are given by the vector fields in the
extended phase space with $t$ treated as the additional coordinate,
where they form the null space of the degenerate two-form
\begin{equation}\label{Sigmaex}
\Sigma=\sigma + dt\wedge d{\cal H}
\end{equation}
with the Hamiltonian given by Eq. (\ref{hvv}) or (\ref{hvV}).
The extended two-form can equivalently be presented in terms of
the integrals of motion being the generators of the extended
Galilei group,
\begin{equation}\label{Sigmaexint}
\Sigma=\frac{1}{m}d{\cal P}_i\wedge d{\cal K}_i
+\frac{1}{2}\theta\epsilon_{ij}d{\cal P}_i\wedge d{\cal P}_j
+\frac{1}{2\kappa}\epsilon_{ij}d{\cal V}_i\wedge d{\cal V}_j.
\end{equation}
The reduction to the surface ${\cal V}_i=0$, $i=1,2$, corresponds
to omitting of the last term in (\ref{Sigmaexint}), and to putting
$V_i=0$ in symplectic form (\ref{sigmaex}) and in Hamiltonian
(\ref{hvV}).

\section{EM-particle: dynamics and symmetries}
Let us consider a charged nonrelativistic particle on the plane
subjected to the external constant homogeneous magnetic $B$ and
electric $E_i$ fields,
\begin{equation}\label{L0}
    L_{EM}=\frac{1}{2}m\dot{x}_i^2 +\frac{1}{2}
B\epsilon_{ij}x_i\dot{x}_j +E_ix_i,
\end{equation}
where we put $c=e=1$, and choose the symmetric gauge,
$A_i(x)=-\frac{1}{2}B\epsilon_{ij}x_j$. By means of the symplectic
two-form
\begin{equation}\label{sigmaEM}
    \sigma=dp_i\wedge dx_i=dP_i\wedge dx_i
+\frac{1}{2}B\epsilon_{ij} dx_i\wedge dx_j,
\end{equation}
the Hamiltonian
\begin{equation}
H=\frac{1}{2m}P_i^2-E_ix_i, \label{hamem}
\end{equation}
generates the equations of motion
\begin{equation}
\frac{d}{dt}x_i=m^{-1}P_i,\qquad \frac{d}{dt}P_i=\omega_c
\epsilon_{ij}(P_j-\omega_c^{-1} \epsilon_{ij}E_j),
 \label{xPE}
\end{equation}
where $P_i=p_i-A_i=p_i+\frac{1}{2}B \epsilon_{ij}x_j$ is a
mechanical momentum, and $\omega_c=\frac{B}{m}$ is the cyclotron
frequency. Comparing (\ref{xPE}) with equations  of motion
(\ref{xvp}) of the NR-anyon in extended formulation, we see a
similarity of the dynamics of  both systems if we identify their
corresponding coordinates $x_i$, treat the dynamical constants $p_i$
of the NR-anyon as analogs of the nondynamical constant quantities
$\omega_c^{-1}\epsilon_{ij}E_j$, and consider the velocities
$m^{-1}P_i$ of the EM-particle,
\begin{equation}\label{PPB}
\{P_i,P_j\}=B \epsilon_{ij},
\end{equation}
 as analogs
of the noncommutative velocities $v_i$  of the NR-anyon. The
analogy between the two systems is extended if we introduce the
variables
\begin{equation}
{\cal X}_i=x_i+B^{-1}\epsilon_{ij}\Pi_j,\qquad
\Pi_i=P_i-\omega_c^{-1}\epsilon_{ij}E_j. \label{XPi}
\end{equation}
The coordinate ${\cal X}_i$ has a sense of the guiding center
coordinate $X_i$ \cite{Ezawa} shifted for the constant electric
term,
\begin{equation}\label{shiftX}
{\cal X}_i=X_i+B^{-1}\omega_c^{-1}E_i,\qquad
X_i=x_i+B^{-1}\epsilon_{ij}P_j.
\end{equation}
The introduced phase space variables are mutually decoupled in the
sense of the Poisson brackets,
\begin{equation}
\{{\cal X}_i,{\cal X}_j\}=-B^{-1}\epsilon_{ij},\qquad \{ {\cal
X}_i,\Pi_j\}=0,\qquad \{\Pi_i,\Pi_j\}=B \epsilon_{ij,}
 \label{XcP}
\end{equation}
allowing us to separate the drift Hall and the circular motions,
\begin{equation}
\frac{d}{dt}{\cal X}_i=B^{-1}\epsilon_{ij}E_j, \qquad
\frac{d}{dt}\Pi_i= \omega_c\epsilon_{ij} \Pi_j. \label{dXc}
\end{equation}
Having in mind the noncommutativity of the ${\cal X}_i$ and their
Poisson bracket decoupling from the $\Pi_i$ variables, we can
treat them, respectively,  as analogs of the covariant NR-anyon
coordinates ${\cal X}_i$ and of the spin variables $V_i$
associated with the Zitterbewegung. The analogy between the
two systems, however, is not complete. The $p_i$ of the NR-anyon
are dynamical and generate translations of the coordinates ${\cal
X}_i$, while the $\omega_c^{-1}\epsilon_{ij}E_j$ are the
nondynamical constants of the EM-particle.

Let us investigate the integrals of motion of the EM-particle and
symmetries associated with them.
 Integration of Eq. (\ref{dXc}) provides us with the four
dynamical integrals of motion
 ($\frac{d}{dt}I=0$,
$\frac{\partial}{\partial t} I\neq 0$):
\begin{equation}
\bar{\cal X}_i={\cal
X}_i-B^{-1}\epsilon_{ij}E_j\cdot t={\cal X}_i(0), \qquad
\bar{\Pi}_i=\Delta_{ij}(t) \Pi_j=\Pi(0), \label{dynint}
\end{equation}
where the matrix $ \Delta_{ij}(t)$ is defined by the equation of
the form (\ref{Del}) with $\omega$ changed for $\omega_c$. These
four independent integrals associated with the drift and circular
motions satisfy the Poisson bracket relations of the form
(\ref{XcP}),
\begin{equation}
\{\bar{\cal X}_i,\bar{\cal X}_j\}=-B^{-1}\epsilon_{ij}, \qquad
\{\bar{\cal X}_i,\bar{\Pi}_j\}=0,\qquad
\{\bar{\Pi}_i,\bar{\Pi}_j\}=B\epsilon_{ij}, \label{barXcP}
\end{equation}
and represent the Hamiltonian (\ref{hamem}) as
\begin{equation}
H=\frac{1}{2m} \left( \bar{\Pi}_i^2+\omega_c^{-2}E_i^2\right)
-E_i\bar{\cal X}_i. \label{HXPi}
\end{equation}
Any other function of $\bar{\cal X}_i$ and $\bar{\Pi}_i$ is also
an integral of motion. In what follows we shall be interested in
the restricted class of the geometric symmetries associated with
the linear and quadratic functions of the integrals
(\ref{dynint}).

{}From Eq. (\ref{dynint}), we find the evolution law
\begin{equation}
x_i(t)=\bar{\cal X}_i+B^{-1}\left(\bar{\Pi}_k\epsilon_{kj}
\Delta_{ji}(t)+\epsilon_{ij}E_j t\right). \label{sEOM}
\end{equation}
Making use of the solutions to the equations of motion, we
calculate the Poisson brackets
\begin{equation}
\{x_i(t_1),x_j(t_2)\}=
-B^{-1}\epsilon_{ik}
\left(
\delta_{kj}-\Delta_{kj}(t_2-t_1)\right),
\label{x1x2}
\end{equation}
\begin{equation}
\{x_i(t_1),p_j(t_2)\}=
\frac{1}{2}
\left(
\delta_{ij}+
\Delta_{ji}(t_2-t_1)\right),
\label{x1p2}
\end{equation}
\begin{equation}
\{x_i(t),\bar{\cal X}_j\}=-B^{-1}\epsilon_{ij},\qquad
\{x_i(t),\bar{\Pi}_j\}=\Delta_{ji}(t). \label{xPi}
\end{equation}

In accordance with Eqs. (\ref{xPi}), (\ref{shiftX}), the rotated
guiding center coordinate
\begin{equation}
{\cal P}_i= -B\epsilon_{ij}\bar{X}_j= \epsilon_{ij}(-B\bar{\cal
X}_j+\omega^{-1}_cE_j)= P_i-B\epsilon_{ij}x_j-E_it \label{transl}
\end{equation}
generates  the translations,
\begin{equation}
\delta x_i=\delta a_i,\qquad \delta t=0, \qquad \delta
L_{EM}=\frac{d}{dt}\left( \delta x_i\left(
\frac{1}{2}B\epsilon_{ij}x_j + eE_it\right) \right),
\label{translt}
\end{equation}
being the Noether charge associated with the displayed
quasi-invariance of the EM-particle Lagrangian. The $\bar{\Pi}_i$
is the Noether charge associated with the symmetry
\begin{equation}
\delta x_i= \delta\lambda_j\Delta_{ji}(t),\quad \delta t=0,\quad
\delta L_{EM}=\frac{d}{dt} \left( \left(
\frac{1}{2}Bx_i-\omega_c^{-1}E_i\right) \epsilon_{ij} \delta
x_j\right), \label{sympi}
\end{equation}
where $\delta\lambda_i$ are the infinitesimal transformation
parameters. Then the linear combination of the integrals
\begin{equation}
{\cal K}_i=\omega_c^{-1}\epsilon_{ij}({\cal R}_j- \bar{\Pi}_j),
\label{Ki}
\end{equation}
with
\begin{equation}
{\cal R}_i={\cal P}_i-\omega_c^{-1}\epsilon_{
ij}E_j=-B\epsilon_{ij}\bar{\cal X}_j, \label{calR}
\end{equation}
can be treated as a generator of the field-deformed Galilei boost
transformations. Indeed, it is the Noether charge associated with
the symmetry
\begin{equation}
\delta x_i=\omega^{-1} \delta v_j\epsilon_{jk}
(\Delta_{ki}(t)-\delta_{ki}), \qquad \delta t=0,\qquad \delta
L_{EM}=\frac{d}{dt} \left(\delta v_i\gamma^b_i\right),
\label{corrGb}
\end{equation}
where
\begin{equation}
\gamma^b_i= \frac{1}{2}m(\delta_{ij}+\Delta_{ij}(t))x_j -
\omega_c^{-2} \epsilon_{ik}(\omega_c t\delta_{kl}
-\epsilon_{kj}\Delta_{jl}(t))E_l. \label{gammab}
\end{equation}
In the limit $B\rightarrow 0$, (\ref{corrGb}) takes the form of
the usual Galilei boost transformations, $\delta x_i\rightarrow
\delta v_i t$, and (\ref{Ki}) is reduced to the boost generator of
a free particle: ${\cal K}_i\rightarrow mx_i-p_it$. On the other
hand, for the usual Galilei boost transformations we have
\begin{equation}
\delta x_i=\delta v_i\cdot t,\quad \delta t=0,\quad \delta
L_{EM}=\frac{d}{dt}\left(\tilde{\gamma}^b_i\delta v_i \right),
\label{GbL}
\end{equation}
with
\begin{equation}
\tilde{\gamma}^b_i=mx_i+\frac{1}{2}B
\epsilon_{ij}x_j\cdot t
+\frac{1}{2}E_i\cdot t^2
-B
\epsilon_{ij}\cdot \int_0^t x_j(\tau)d\tau.
\label{Gbg}
\end{equation}
Due to the last term in Eq. (\ref{Gbg}), the Noether charge
associated with Galilei boosts (\ref{GbL}),
\begin{equation}
\tilde{\cal K}_i=\tilde{\gamma}^b_i-tp_i,
\label{jb}
\end{equation}
is nonlocal in time. By means of relations (\ref{x1x2}) and
(\ref{x1p2}), we find that the nonlocal in time Noether charge
(\ref{jb}) generates not the transformation (\ref{GbL}), (\ref{Gbg})
but the field-deformed boost transformations (\ref{corrGb}),
(\ref{gammab}) associated with the local in time dynamical integral
of motion (\ref{Ki}). This happens because to find the
transformations generated by (\ref{jb}), we have to use the Poisson
brackets (\ref{x1x2}) calculated on shell, i.e. with making use of
the solutions to the equations of motion. The calculation of the
integral in the last term of (\ref{Gbg}) with the help of  solution
(\ref{sEOM}) reduces the nonlocal charge (\ref{jb}) into the local
integral of motion (\ref{Ki}).

Taking into account the picture of transmutation on shell of the
nonlocal in time Noether charges  related to the Galilei boosts into
the local in time field-deformed Galilei symmetry generators, let us
analyze in the same way the rotation transformations. For
infinitesimal rotations we have
\begin{equation}
\delta x_i=\delta \varphi\cdot \epsilon_{ij}x_j,\quad \delta
t=0,\quad \delta L_{EM}=\frac{d}{dt}\left(\gamma^r\delta
\varphi\right), \label{rot}
\end{equation}
with
$$
 \gamma^r=\epsilon_{ij}E_i\int_0^t x_j(\tau)d\tau.
$$
As in the previous case, the associated Noether charge
$j^r=\epsilon_{ij}x_ip_j+\gamma^r$ is not local in time quantity.
On shell, it transmutes into the local in time integral
\begin{equation}
{\cal J}=\frac{1}{2}\left( B\bar{X}_i{}^2-B^{-1}
\bar{P}_i^2\right), \label{tJ}
\end{equation}
where
$\bar{P}_i=\bar{\Pi}_i+\omega_c^{-1}\epsilon_{ij}E_j=P_i(0)$.
 The integral (\ref{tJ}) generates the field-deformed
rotation transformations
\begin{equation}
\delta x_i= \{{\cal J},x_i\} \delta \varphi= \epsilon_{ij}\left(
x_j-m^{-1}\omega_c^{-2}\left(E_j\cdot (1-\cos\omega_c t)-
\epsilon_{jk} E_k\cdot (\omega_c t-\sin\omega_c t)\right)
\right)\cdot \delta\varphi, \label{rotimp}
\end{equation}
which at $E_i=0$ are reduced to the usual rotation
transformations, while the integral (\ref{tJ}) takes a usual form
${\cal J}=\epsilon_{ij}x_ip_j$.

Note that for $B\rightarrow 0$,
Eq. (\ref{rotimp}) gives $\delta x_i=
\epsilon_{ij}(x_j-\frac{1}{2} m^{-1}E_j\cdot t^2)\delta\varphi$.
Hence, in the limit $B\rightarrow 0$, the infinitesimal
translations, the field-deformed Galilei boosts and the rotation
transformations are polynomials of $t$ of the degree 0, 1 and 2,
respectively.

The integrals ${\cal P}_i$, ${\cal K}_i$, ${\cal J}$ and $H$ form
the field-deformed Galilei algebra
\begin{eqnarray}
&\{{\cal P}_i,{\cal P}_j\}= -B\epsilon_{ij},\qquad \{{\cal
K}_i,{\cal K}_j\}=0,\qquad \{{\cal K}_i,{\cal
P}_j\}=m\delta_{ij},&\nonumber\\
&\{{\cal J},{\cal P}_i\}=\epsilon_{ij} {\cal P}_j,\qquad \{{\cal
J},{\cal K}_i\}=\epsilon_{ij} {\cal K}_j,& \label{JPK}\\
&\{{\cal P}_i,H\}=E_i,\qquad \{{\cal K}_i,H\}={\cal
P}_i+\omega_c\epsilon_{ij}{\cal K}_j, \qquad \{{\cal
J},H\}=-m^{-1}\epsilon_{ij}E_i{\cal K}_j.&\nonumber
\end{eqnarray}
The field-deformed Galilei algebra (\ref{JPK}) has the two Casimir
elements
\begin{equation}\label{J=PK}
{\cal C}_1= \epsilon_{ij}{\cal K}_i{\cal P}_j
-\frac{1}{2}\omega_c{\cal K}_i^2 - m{\cal J},\qquad
 {\cal C}_2=
{\cal P}_i^2- 2B {\cal J} -2{\cal K}_iE_i -2mH,
\end{equation}
both of which take zero values in the EM-particle model.
 Note also that
$ {\cal P}_i=p_i(0)-\frac{1}{2}B \epsilon_{ij}x_j(0),$ $ {\cal
K}_i=mx_i(0)$ and $ {\cal J}=\epsilon_{ij}x_i(0)p_j(0). $

\section{Extended model and enlarged exotic Galilei symmetry}

We have seen that the Noether charges associated with the Galilei
boosts and rotations of the EM-particle are nonlocal in time. On
shell, they are transformed into the local integrals generating
field-deformed symmetry transformations of a complicated form
generalizing the initial boost and rotation transformations. Let
us look for such an extension of the EM-particle model which
\begin{itemize}
    \item would not change its dynamics,
    \item but would eliminate the
nonlocality of the boost and rotation Noether charges.
\end{itemize}
Since the nonlocality of the both charges originates from the time
integral of $x_i(t)$, the work would be done if we present  the
particle coordinates as a time derivative of some ``pre-coordinate''
variables $q_i$. Taking  also into account that more profound
analogy between the EM-particle and NR-anyon is destroyed by the
non-dynamical character of the electric field, we change the
nondynamical quantities $E_i$ for the dynamical ones ${\cal E}_i$,
and add to the Lagrangian the term ${\cal E}_i\dot{q}_i$, i.e.
change (\ref{L0}) for the extended Lagrangian system
\begin{equation}
L_{ext}=\frac{1}{2}m\dot{x}_i^2+ \frac{1}{2}B\epsilon_{ij}x_i\dot{x}_j
+{\cal E}_ix_i+{\cal E}_i\dot{q}_i, \label{Lext1}
\end{equation}
treating the ${\cal E}_i$ as the momenta for the new coordinates
$q_i$. Since the equations of motion for $q_i$ read
\begin{equation}
\label{Edot}
\dot{\cal E}_i=0,
\end{equation}
the $E_i$ transmutes here into the dynamical integral of motion
${\cal E}_i$. The additional term does not change equations of
motion for $x_i$ (\ref{xPE}), but provides us with the new desired
element: on l.h.s. of Eq. (\ref{xPE}) the
$\omega_c^{-1}\epsilon_{ij}E_j$ is changed now for the dynamical
constant $\omega_c^{-1}\epsilon_{ij}{\cal E}_j$ to be similar to the
$p_i$ in the NR-anyon equations of motion (\ref{xvp}). The two last
terms in (\ref{Lext1}) mean that the ${\cal E}_i$ plays
simultaneously the role of the Lagrange multiplier generating the
constraint
\begin{equation}
\label{xq0}
x_i+\dot{q}_i=0.
\end{equation}
Taking into account this constraint,
the nonlocal term $\int^t x_i(\tau)d\tau$ in the boost and
rotation Noether charges of the EM-particle is reduced to the
local term $-q_i(t)$.

In the Hamiltonian picture the extended system (\ref{Lext1}) is
described by the symplectic two-form
\begin{equation}\label{sigma1}
    \sigma=dp_i\wedge dx_i + d{\cal E}_i\wedge dq_i= dP_i\wedge dx_i
+\frac{1}{2}B\epsilon_{ij} dx_i\wedge dx_j + d{\cal E}_i\wedge
dq_i.
\end{equation}
The Hamiltonian of (\ref{Lext1}) has the same form (\ref{hamem})
with the $E_i$ changed for the dynamical constant ${\cal E}_i$,
while the Hamiltonian equations of motion (\ref{xPE}) are
supplemented with the equations (\ref{Edot}) and
(\ref{xq0}). Making use of the equations of motion, we find
the complete set of dynamical integrals of motion linear in the
extended phase space variables,
\begin{equation}
{\cal P}_i,\qquad {\cal K}_i=mx_i +B\epsilon_{ij}q_j -{\cal
P}_it-\frac{1}{2}{\cal E}_it^2,\qquad \bar{\Pi}_i, \qquad {\cal
E}_i, \label{PKPE}
\end{equation}
where ${\cal P}_i$ and $\bar{\Pi}_i$ are given by the equations of
the form (\ref{transl}), (\ref{dynint}) with the $E_i$ changed for
the dynamical ${\cal E}_i$. We have also the  quadratic
integrals\footnote{We are not interested here in other quadratic
integrals of motion, in particular, in those which could be
related to the generalized dilatation and special conformal
symmetries.}
\begin{equation}
{\cal H}=\frac{1}{2m}P_i^2-{\cal E}_ix_i, \label{Hamex}
\end{equation}
\begin{equation}
{\cal J}=\epsilon_{ij}(x_ip_j+q_i{\cal E}_j)=
\epsilon_{ij}(x_iP_j+q_i{\cal E}_j)+ \frac{1}{2}Bx_i^2,
\label{Jex}
\end{equation}
being the generators of the time translations and usual (not
deformed) rotations.

Lagrangian (\ref{Lext1}) gives, however, not the most general
extension of the system with the described properties. Indeed, the
Chern-Simons-like term $\frac{1}{2}\beta\epsilon_{ij}{\cal
E}_i\dot{\cal E}_j$ with a parameter $\beta$ of dimensionality
$B^{-1}\omega_c^{-2}$ can be added without changing the equations
of motion. Its only effect will be to modify the Poisson brackets
of $q_i$ making them to be non-commuting coordinates. Summarizing,
we arrive at the following Lagrangian for the extended exotic system
\begin{equation}\label{LCen}
    L_{Ex}=\frac{1}{2}m\dot{x}_i^2+ \frac{1}{2}B\epsilon_{ij}x_i\dot{x}_j
+{\cal E}_ix_i+{\cal E}_i\dot{q}_i
+\frac{1}{2}\beta\epsilon_{ij}{\cal E}_i\dot{\cal E}_j.
\end{equation}
If we switch off the electromagnetic interaction by restoring the
electric charge $e$ and then put $e=0$, the
system (\ref{LCen}) is transformed into the decoupled sum of the
two free nonrelativistic particles --- the usual one of mass $m$,
and the exotic particle with noncommuting coordinates \cite{DH} of
zero mass.

In canonical formalism, the system (\ref{LCen}) is given by the
nontrivial Poisson brackets
\begin{equation}\label{PBex}
    \{x_i,P_j\}=\delta_{ij},\qquad
    \{P_i,P_j\}=B\epsilon_{ij},\qquad
    \{q_i,q_j\}=\beta\epsilon_{ij},\qquad
    \{q_i,{\cal E}_j\}=\delta_{ij},
\end{equation}
and by the Hamiltonian (\ref{Hamex}). The complete set of
equations of motion is (\ref{Edot}), (\ref{xq0}) and
\begin{equation}
\dot{x}_i=\frac{1}{m}P_i,\qquad \dot
P_i=\omega_c\epsilon_{ij}(P_j-\omega_c^{-1} \epsilon_{jk}{\cal
E}_k).
\label{eomext}
\end{equation}
The quantities (\ref{PKPE}), (\ref{Hamex}) and (\ref{Jex}) are
the integrals of motion of the system (\ref{LCen}) being the
Noether charges generating  the following symmetries:
\begin{eqnarray}
\begin{array}{cccc}
  {\cal P}_i: & \delta x_i=\delta a_i, & \delta q_i=-t\delta a_i, & \delta {\cal E}_i=0, \\
  {\cal K}_i: & \delta x_i=t\delta v_i, & \delta q_i=-\left(\frac{1}{2}t^2+\beta B\right)\delta v_i,
  & \delta {\cal E}_i=-B\epsilon_{ij}\delta v_j, \\
  {\cal E}_i: & \delta x_i=0, & \delta
q_i=\delta d_i, & \delta {\cal E}_j=0, \\
  \bar{\Pi}_i: & \delta x_i= \delta \lambda_j\Delta_{ji}(t), & \delta
q_i=-\omega_c^{-1}\delta\lambda_j \epsilon_{jk}\Delta_{ki}(t), & \delta {\cal E}_i=0, \\
  {\cal J}: & \delta x_i=\delta\varphi \,\epsilon_{ij}x_j, & \delta
q_i=\delta\varphi \,\epsilon_{ij}q_j, & \delta {\cal
E}_i=\delta\varphi \,\epsilon_{ij}{\cal E}_j,\\
\end{array}
 \label{transext}
\end{eqnarray}
where $\Delta_{ij}(t)$ is defined by Eq. (\ref{Del}), while
$\delta a_i$, $\delta v_i$, $\delta d_i$, $\delta \lambda_i$ and
$\delta\varphi$ are the transformation parameters.
 The Hamiltonian (\ref{Hamex})
generating the time translations can be presented in the
equivalent form in terms of other integrals of motion,
\begin{equation}\label{HCen}
    {\cal H}=\frac{1}{2m}(\bar{\Pi}_i^2+\omega_c^{-2}{\cal
E}_i^2)-\frac{1}{B}\epsilon_{ij}{\cal E}_i{\cal R}_j,
\end{equation}
where the ${\cal R}_i$ is given by the equation of the form
(\ref{calR}) with the $E_i$ changed for ${\cal E}_i$.

The integrals (\ref{PKPE}), (\ref{Hamex}) and (\ref{Jex}) form the
enlarged exotic Galilei algebra,
\begin{eqnarray}
&\{{\cal P}_i,{\cal P}_j\}=-B\epsilon_{ij},\qquad \{{\cal
K}_i,{\cal P}_j\}=m\delta_{ij},\qquad
 \{{\cal K}_i,{\cal
K}_j\}=\beta B^2\epsilon_{ij},\qquad \{{\cal K}_i,{\cal
E}_j\}=B\epsilon_{ij},&
\nonumber
\\
&\{\bar{\Pi}_i,\bar{\Pi}_j\}=B\epsilon_{ij},\qquad \{{\cal
J},{\cal Q}_i\}=\epsilon_{ij}{\cal Q}_j,\qquad {\cal Q}_i={\cal
P}_i,\, {\cal K}_i,\, \bar{\Pi}_i,\, {\cal E}_i,&\label{jpk}
\\
&\{{\cal H},{\cal K}_i\}=-{\cal P}_i,\qquad \{{\cal H},{\cal
P}_i\}=-{\cal E}_i,\qquad \{{\cal
H},\bar{\Pi}_i\}=-\omega_c\epsilon_{ij}\bar{\Pi}_j,&\nonumber
\end{eqnarray}
where the nontrivial brackets are displayed only,  cf.
\cite{NO,NOT}. The algebra (\ref{jpk}) is characterized by the two
Casimir elements
\begin{equation}\label{CC12}
    {\cal C}_1={\cal R}^2_i-\bar{\Pi}_i^2-2{\cal E}_i{\cal K}_i+
    \beta B
    {\cal E}_i^2-2B{\cal J},\qquad
    {\cal C}_2=\bar{\Pi}_i^2-2\omega_c^{-1}\epsilon_{ij}{\cal E}_i{\cal P}_j-
    \omega_c^{-2}{\cal E}_i^2-2m{\cal H},
\end{equation}
which for system (\ref{LCen}) take zero values.

The coordinates ${\cal X}_i$ defined  by Eq. (\ref{XPi}) with the
$E_i$ changed for the ${\cal E}_i$, i.e.
$$
{\cal X}_i=x_i+\frac{1}{B}\epsilon_{ij}
\left(P_j-\omega_c^{-1}\epsilon_{jk}{\cal
E}_k\right)=\frac{1}{B}\epsilon_{ij}\left({\cal R}_j +{\cal E}_j
t\right),
$$
can be identified as analogs of covariant non-commuting
coordinates (\ref{XxV}) of the NR-anyon, $\{{\cal X}_i,{\cal
X}_j\}=-B^{-1}\epsilon_{ij}$. Like the $x_i$, they satisfy the
relations $\{{\cal P}_i,{\cal X}_j\}=-\delta_{ij}$, $\{{\cal
K}_i,{\cal X}_j\}=t\,\delta_{ij}$, $\{{\cal E}_i,{\cal X}_j\}=0$,
but, unlike the latter, $\{\bar{\Pi}_i,{\cal X}_j\}=0$. As a
consequence,  in correspondence with the structure of the
Hamiltonian (\ref{HCen}), the evolution of the ${\cal X}_i$
reveals only the translation Hall drift motion, $\frac{d}{dt}{\cal
X}_i=\frac{1}{B}\epsilon_{ij}{\cal E}_j$.

\section{Reduction of the \emph{Ex}-model to the NR-anyon and EM-particle}

The translation generators ${\cal P}_i$, $i=1,2$,  of the {\it Ex}-model
do not commute in contrast with those of the free NR-anyon, while
its boost generators ${\cal K}_i$, $i=1,2$, unlike the EM-particle,
do not commute at $\beta\neq 0$. Nevertheless, we shall show below
that both systems, the EM-particle and NR-anyon, are contained in
the {\it Ex}-model in the form of subsystems, and could be ``extracted"
from the latter by the appropriate Hamiltonian reductions.

First, consider the reduction which produces the NR-anyon model. To
this end, we define
\begin{equation}\label{PNRH}
{\cal P}^{NR}_i=\omega_c^{-1}\epsilon_{ij}{\cal E}_j,\qquad {\cal
H}^{NR}={\cal H}-\frac{1}{m}{\cal P}^{NR}_i{\cal
R}_i=\frac{1}{2m}\left(\bar{\Pi}_i^2+({\cal P}^{NR}_i)^2\right).
\end{equation}
Identifying the parameter $\kappa$ of the NR-anyon with
$-m^{2}B^{-1}$ of the {\it Ex}-model, and then fixing in the latter
model $\beta =-m^2B^{-3}$, we find that the set of integrals of
motion ${\cal P}^{NR}_i$, $\bar{\Pi}_i$, ${\cal K}_i$, ${\cal
H}^{NR}$ and ${\cal J}$ of the \emph{Ex}-model forms the
algebra like that formed by the integrals ${\cal P}_i$,
$\omega{\cal V}_i$, ${\cal
K}_i$, ${\cal H}$ and ${\cal J}$ of the NR-anyon. The only
difference is the sign of the brackets of the $\omega{\cal V}_i$
with $\omega{\cal V}_j$ (see Eq. (\ref{GalV})) in comparison with
the brackets of $\bar{\Pi}_i$ with $\bar{\Pi}_j$  in (\ref{jpk}).
Correspondingly, the sign before the term
$\bar{\Pi}_i^2$ in the ${\cal H}^{NR}$ is different from the sign
before the ${\omega^2}{\cal V}_i^2$ in the Hamiltonian (\ref{hvV})
of the extended formulation for the NR-anyon. At the same time, note
that the integral ${\cal R}_i={\cal P}_i-{\cal P}^{NR}_i$ has
vanishing Poisson brackets with the integrals ${\cal P}^{NR}_i$,
$\bar{\Pi}_i$, ${\cal K}_i$, ${\cal H}^{NR}$ of the \emph{Ex}-model,
and is rotated by the ${\cal J}$.

The {\it Ex}-model coordinates, which are covariant with respect to the
action of the integrals ${\cal P}^{NR}_i$, $\bar{\Pi}_i$, ${\cal
K}_i$, ${\cal H}^{NR}$ and ${\cal J}$, can be identified in two
steps. First, analogously to the NR-anyon, we present the boost
generator from (\ref{PKPE}) in the form ${\cal K}_i=m{\cal
Y}_i^{NR}-{\cal P}^{NR}_i t$, cf. Eq. (\ref{KJY}). This gives
\begin{equation}\label{YNR}
    {\cal Y}_i^{NR}=x_i+\omega_c\epsilon_{ij}q_j-\frac{1}{2m}{\cal
    E}_it^2-\frac{1}{m}{\cal R}_it.
\end{equation}
Then the
\begin{equation}\label{XCen}
{\cal X}^{NR}_i={\cal Y}^{NR}_i+\beta\omega_c^2\epsilon_{ij}{\cal
P}_j^{NR}= x_i+\omega_c\epsilon_{ij}q_j-\beta\omega_c {\cal E}_i
-\frac{1}{m}{\cal R}_it-\frac{1}{2m}{\cal E}_it^2
\end{equation}
(cf. Eq. (\ref{Y}))
 is the analog of the covariant noncommuting
coordinates of the NR-anyon in the extended formulation,
$$
\{{\cal X}^{NR}_i,{\cal
X}^{NR}_j\}=-\beta\omega_c^2\epsilon_{ij},\qquad \{{\cal
X}^{NR}_i,{\cal Y}^{NR}_j\}=0,\qquad \{{\cal X}^{NR}_i,{\cal
P}^{NR}_j\}=\delta_{ij},
$$
$$
 \{{\cal
K}_i,{\cal X}^{NR}_j\}=\delta_{ij}t,\qquad
 \{\bar{\Pi}_i,{\cal
X}^{NR}_j\}=0,\qquad \{{\cal J},{\cal
X}^{NR}_i\}=\epsilon_{ij}{\cal X}^{NR}_j.
$$
The ${\cal X}^{NR}_i$,
${\cal Y}^{NR}_i$ and $\bar{\Pi}_i$ diagonalize the rotation
generator,
$$
{\cal J}=\frac{1}{2\beta\omega_c^2}\left( ({\cal
X}^{NR}_i)^2-({\cal Y}^{NR}_i)^2\right)
-\frac{1}{2B}\bar{\Pi}_i^2.
$$

Therefore, the
{\it Ex}-model being reduced to the surface of the two second
 class constraints
\begin{equation}\label{conR}
    {\cal R}_i={\cal P}_i-{\cal P}^{NR}_i\approx 0
\end{equation}
is transformed into the NR-anyon \cite{HP2,HP3} (with the only sign
difference mentioned above). In the sense of the Dirac theory of the
constrained systems, in correspondence with definition (\ref{A*})
taken for the second class constraints (\ref{conR}), we have ${\cal
P}_i^*={\cal P}_i^{NR}$, ${\cal K}_i^*={\cal K}_i$,
$\bar{\Pi}_i^*=\bar{\Pi}_i$, ${\cal H}^*={\cal H}^{NR}$. The first
from these relations means that in the subspace defined by the
constraints (\ref{conR}),  the vector ${\cal P}^{NR}_i$ takes the
role of the translation generator ${\cal P}_i$, i.e. the reduction
to the surface (\ref{conR}) provokes a kind of {\it
electric-translation} transmutation.

 Further reduction to the surface of zero values of the
circular motion integrals,
\begin{equation}\label{Pi0}
\bar{\Pi}_i\approx 0,
\end{equation}
``extracts''  from the extended system the model \cite{DH} of a free
nonrelativistic anyon in the minimal formulation.

Let us consider in more detail the reduction  procedure of the
system to the subspaces of the second class constraints given in
terms of integrals of motion. The symplectic two-form
corresponding to Poisson brackets (\ref{PBex}) is
\begin{equation}
    \sigma=dP_i\wedge dx_i +\frac{1}{2}B\epsilon_{ij}\, dx_i\wedge
    dx_j+d{\cal E}_i\wedge dq_i +\frac{1}{2}\beta\epsilon_{ij}\, d{\cal E}_i\wedge
    d{\cal E}_j,\label{sigext}
\end{equation}
In terms of the variables ${\cal X}_i$ and $\Pi_i$ given by Eq.
(\ref{XPi}) (in which the $E_i$ is changed for ${\cal E}_i$), it
takes the form
\begin{equation}
    \sigma=\frac{1}{2}\epsilon_{ij}\left(
    B\, d{\cal X}_i\wedge d{\cal X}_j-B^{-1}d\Pi_i\wedge
    d\Pi_j+\beta\, d{\cal E}_i\wedge d{\cal E}_j\right)+
        d{\cal E}_i\wedge dq^x_i,\label{sigext1}
\end{equation}
where
\begin{equation}\label{Q}
   q^x_i =q_i-\omega_c^{-1}\epsilon_{ij}x_j.
\end{equation}
I.e., the separation of the circular and drift motions is
accompanied here  by the appearance of the variables $q^x_i$
mixing the noncommuting coordinates $q_i$ with the commuting ones
$x_i$. Since
\begin{equation}\label{redin}
    \frac{1}{2B}\epsilon_{ij}\, d\Pi_i\wedge
    d\Pi_j=\frac{1}{2B}\epsilon_{ij}\,
    d\bar{\Pi}_i\wedge d\bar{\Pi}_j-\frac{1}{2m}dt\wedge
    d\bar{\Pi}_i^2,
\end{equation}
the transition from the variable $\Pi_i$ to the associated
integrals $\bar{\Pi}_i$ is accompanied by the shift of the
Hamiltonian for the term $-\frac{1}{2m}\bar{\Pi}_i^2$. Having in
mind the relation $\bar{\cal X}_i=B^{-1}\epsilon_{ij}{\cal R}_j$,
we get
\begin{equation}\label{XRred}
\frac{1}{2}B\epsilon_{ij}\,d{\cal X}_i\wedge d{\cal X}_j+d{\cal
E}_i\wedge dq^x_i=\frac{1}{2}B^{-1}\epsilon_{ij}\, d{\cal
R}_i\wedge d{\cal R}_j +d{\cal E}_i\wedge dQ_i+dt\wedge d({\cal
E}_i\bar{\cal X}_i),
\end{equation}
where
\begin{equation}\label{Qqx}
Q_i=q^x_i+\frac{1}{B}\epsilon_{ij}\left(t{\cal R}_j +\frac{1}{2}\,
{\cal E}_j t^2\right).
\end{equation}
Therefore, the transition from the ${\cal X}_i$ to the associated
integrals ${\cal R}_i=-B\epsilon_{ij}\bar{\cal X}_j$ is
accompanied by the shift of the Hamiltonian for the term ${\cal
E}_i\bar{\cal X}_i={\cal E}_i{\cal X}_i$, and by the transition
from the variables $q^x_i$ to the variables $Q_i$, which, like the
$q_i$, are the classically noncommuting variables, $
    \{Q_i,Q_j\}=\beta\epsilon_{ij}.
$
Finally, the relation
\begin{equation}\label{KE}
d{\cal E}_i\wedge dQ_i=-B^{-1}\epsilon_{ij}\, d{\cal E}_i\wedge
d{\cal K}_j +dt\wedge d\left(-\frac{1}{2m}\omega_c^{-2}{\cal
E}_i^2\right)
\end{equation}
means that the transition from the $Q_i$ to the integral ${\cal
K}_i$ is accompanied by the shift of the Hamiltonian for the term
$-\frac{1}{2m}\omega_c^{-2}{\cal E}_i^2$. The three steps can be
summarized in presentation of the extended two-form
$\Sigma=\sigma+dt\wedge d{\cal H}$ in terms of the integrals of
motion,
\begin{equation}
    \Sigma=\frac{1}{2B}\epsilon_{ij}
    \left(d{\cal R}_i\wedge d{\cal R}_j-d\bar{\Pi}_i\wedge
    d\bar{\Pi}_j\right)
    -\frac{1}{B}\epsilon_{ij}\, d{\cal E}_i\wedge d{\cal K}_j+
    \frac{1}{2}\beta \epsilon_{ij}\, d{\cal E}_i\wedge d{\cal
    E}_j,
    \label{Sigex}
\end{equation}
i.e. the three described shifts cancel all the three corresponding
terms in Hamiltonian (\ref{Hamex}) reducing it to zero. This is in
accordance  with the nature of the independent variables ${\cal
R}_i$, $\bar{\Pi}_i$, ${\cal K}_i$ and ${\cal E}_i$, which are the
integrals of motion, and can be used instead of the initial
variables $x_i$, $P_i$, $q_i$ and ${\cal E}_i$ for the canonical
description of the {\it Ex}-model. For the sake of completeness, note also
that with respect to the set of the second class constraints
(\ref{conR}) and (\ref{Pi0}) we have $x_i^{*}=\frac{1}{m}{\cal
P}^{NR}_it$ and $(q^x_i)^*=q^x_i-\frac{1}{B}\epsilon_{ij}{\cal R}_j
t$.

{}From (\ref{Sigex}) it follows that the vector ${\cal P}_i^{NR}$
with commuting components is
conjugate to the boost generators in the same way as a translation
generator, $\{{\cal K}_i,{\cal P}^{NR}_j\}=m\delta_{ij}$, cf. the
vector $p_i$ of the NR-anyon. The noncovariant coordinate
(\ref{YNR}) is nothing else as the rotated vector (\ref{Qqx}), $
{\cal Y}^{NR}_i=\omega_c\epsilon_{ij}Q_j$. Note also that the
presence of the last term in (\ref{Qqx}) gives $\frac{\partial
Q_i}{\partial t}=\frac{1}{B}\epsilon_{ij}{\cal R}_j$, that  turns
into zero on the surface (\ref{conR}). As a consequence, on the
reduced phase space given by the second class constraints
(\ref{conR}), (\ref{Pi0}), the covariant coordinates ${\cal
X}^{NR}_i$ do not depend explicitly on evolution parameter $t$,
and the system admits the Hamiltonian description in terms of the
extended two-form of the NR-anyon model in minimal formulation,
$$
\Sigma^{NR}_0=\sigma+dt\wedge d{\cal H}_{0},\qquad \sigma= d{\cal
P}^{NR}_i\wedge d{\cal X}^{NR}_i+ \theta d{\cal P}^{NR}_i\wedge
d{\cal P}^{NR}_i,\qquad {\cal H}_0=\frac{1}{2m}\left({\cal
P}_{i}^{NR}\right)^2,
$$
 where we have taken into account the
described above identification of the parameters of the {\it Ex}-model
with those of the NR-anyon.

The phase space reduction of the {\it Ex}-model to the EM-particle is the
reverse of the procedure with which the extended model was obtained
from the EM-particle. To realize it, we have to fix the dynamical
variables ${\cal E}_i$ to be equal to the constant electric field
components $E_i$, as well as to eliminate the variables $q_i$,
conjugate to the momenta ${\cal E}_i$, by presenting them in terms
of other variables. This should be realized in a way consistent
with the dynamics of the {\it Ex}-model. To satisfy the requirement
of consistency,
and having in mind the EM-particle relation (\ref{Ki}), we introduce
the four second class constraints
\begin{equation}\label{red1}
    {\cal E}_i-E_i\approx 0,\qquad
    {\cal K}_i-\omega_c^{-1}\epsilon_{ij}({\cal R}_j-
    \bar{\Pi}_j)\approx 0.
\end{equation}
Being given in terms of the integrals of motion, the subspace
(\ref{red1}) is stable under the system evolution. Since the ${\cal
E}_i$ has nontrivial brackets with the $q_i$ only, the reduction to
the surface of the constraints (\ref{red1}) means effectively the
elimination of the $q_i$ (by presenting it in terms of other phase
space variables and time) and reduction of the dynamical ${\cal
E}_i$ to the nondynamical electric field $E_i$. The {\it Ex}-model
symplectic two-form (\ref{sigext}) is reduced on the surface
(\ref{red1}) to the symplectic form (\ref{sigmaEM}) of the
EM-particle. Relation (\ref{KE}) corresponds to the elimination from
the Hamiltonian of the term $\frac{1}{2m}\omega_c^{-2}{\cal E}_i^2$,
which is the kinetic term for the noncommuting variables $q_i$. As a
result, the Dirac brackets of the remaining independent phase space
variables $x_i$ and $P_i$ calculated with the second class
constraints (\ref{red1}) coincide with the Poisson brackets of the
EM-particle, while the Hamiltonian ${\cal H}$, the boost and
rotation generators take the form of the corresponding integrals of
the EM-particle given by Eqs. (\ref{hamem}), (\ref{Ki}) and
(\ref{tJ}). Since the integral ${\cal P}_i$ commutes with the
constraints (\ref{red1}), its Dirac brackets with any other quantity
coincide with the initial brackets. This means, in particular, that
after reduction the translation symmetry  has a usual form, $\{{\cal
P}_i, x_j\}^*=-\delta_{ij}$. On the other hand, the brackets for
${\cal K}_i$ and ${\cal J}$ with $x_j$  change their usual form
corresponding to transformations (\ref{transext}) for that
corresponding to the field-deformed transformation laws
(\ref{corrGb}) and (\ref{rotimp}) of the EM-particle.

Therefore, the {\it Ex}-model reduced to the surface of
the constraints (\ref{red1}) is the EM-particle model. The usual
Galilei boost and rotation transformation symmetries of the extended
system transmute under this reduction into the corresponding
field-deformed transformation symmetries of
the EM-particle model, while the Dirac bracket relations of ${\cal
P}_i$, ${\cal K}_i$, ${\cal J}$ and ${\cal H}$ take the form of the
Poisson bracket relations (\ref{JPK}).

\section{Discussion and concluding remarks}

We have considered two schemes of reduction of the {\it Ex}-model to the
surfaces given by the sets of second class constraints (\ref{conR})
and (\ref{Pi0}),  and (\ref{red1}). The surfaces are presented in
terms of the integrals of motion of the {\it Ex}-model, that guarantees
the consistency of the reduction procedure with the model dynamics.
In the first reduction scheme, the boost generators of the {\it Ex}-model
commute with the constraints (\ref{conR}), (\ref{Pi0}), while in the
second scheme its translation generators commute with constraints
(\ref{red1}). As a result, we ``extract" from the extended system,
correspondingly, the NR-anyon and the EM-particle models.

Note that in a generic case not any surface given by the dynamical
integrals of motion in the form of second class constraints can be
used for a reduction. For instance, the second pair of relations
from (\ref{red1}) forms  alone a subset of two second class
constraints. However, the reduction of the extended two-form
(\ref{Sigex}) to the surface of such constraints can not be
presented in the form $\Sigma_{red}=\sigma_{red}+dt\wedge {\cal
H}_{red}$ with non-degenerate two-form $\sigma_{red}$ and reduced
Hamiltonian ${\cal H}_{red}$, where the symplectic  matrix
associated with $\sigma_{red}$ would not depend explicitly on time.
This means that after reduction to such a surface we can not
describe the resulting system as a Hamiltonian one.

On the other hand, the two second class constraints (\ref{Pi0})
can be used for the Hamiltonian reduction. It  produces the system
described by the noncommuting coordinates
$$
{\cal X}_i=x_i^*=x_i+{B}^{-1}\epsilon_{ij}{\Pi}_j\approx
x_i,\qquad \{{\cal X}_i,{\cal X}_j\}=-{B}^{-1}\epsilon_{ij},
$$
and by the decoupled in the sense of Dirac brackets noncommuting
coordinates $q^x_i$ and conjugate momenta ${\cal E}_i$,
$$
\{q^x_i,q^x_j\}=\beta\epsilon_{ij},\qquad \{q^x_i,{\cal
E}_j\}=\delta_{ij}.
$$
The Hamiltonian of such a reduced system is
$$
{\cal H}^H_{ext}=\frac{1}{2m}\omega_c^{-2}{\cal E}_i^2-{\cal
E}_i{\cal X}_i.
$$
 It generates  the evolution corresponding to the
Hall motion, $\frac{d}{dt}{\cal X}_i=\frac{1}{B}\epsilon_{ij}{\cal
E}_j$, as well as the equations
$\frac{d}{dt}{q^x_i}=m^{-1}\omega_c^{-2}{\cal E}_i-{\cal X}_i$,
$\frac{d}{dt}{\cal E}_i=0.$ This system possesses the symmetry
corresponding to the enlarged exotic Galilei algebra (\ref{jpk})
with the omitted bracket relations which include the $\bar{\Pi}_i$
and turn into zero after reduction to the surface
(\ref{Pi0})\footnote{Therefore, the earlier discussed
Galilei-Maxwell symmetry \cite{NO,NOT} corresponds to the reduced
Hall motion rather than to the generic motion of the charged
particle.}. The system can be further reduced to the surface given
by Eq. (\ref{red1}), and we obtain the system given in terms of the
noncommuting coordinates ${\cal X}_i$ and the Hamiltonian ${\cal
H}_{Hall}=-E_i{\cal X}_i$, which describes the Hall motion.

After reduction to the surface of the subsystem of the second
class constraints (\ref{conR}), we obtain the system
equivalent to the NR-anyon in extended formulation, which includes
the analogs of the anyon  noncommuting velocities. The only
difference is in the signs of the Poisson brackets of the
integrals $\bar{\Pi}_i$ in comparison with those of the NR-anyon
integrals ${\cal V}_i$, and in the signs of the corresponding
quadratic terms in  Hamiltonians (\ref{hvV}) and (\ref{PNRH}). As
a result, the Hamiltonian (\ref{PNRH}) is positively definite,
while the Hamiltonian (\ref{hvV}), appearing in the
higher-derivative model for the noncommutative plane \cite{LSZ},
is not. This difference turns out to be essential at the quantum
level (see the discussion in \cite{HP1,HP2}).

In the same vein as we constructed the extended system proceeding
from the EM-particle model, one can start from the NR-anyon in a
background of the constant electric and magnetic fields
\cite{DH,HP3}. In such a case from the very beginning we shall get
the noncommuting translations and noncommuting, in a generic case,
Galilei boosts. By the appropriate choice of the multiplicative
parameter in the electric Chern-Simons term, the Galilei boosts of
such an extended system can be made to be commuting (cf. the
enlarged exotic Galilei symmetry of the \emph{Ex}-model at
$\beta=0$). Note here that some analogous extension of the exotic
particle \cite{DH} in background homogeneous electromagnetic field
was discussed recently in \cite{HMS}. The essential difference of
the construction \cite{HMS} is that there the components of electric
field were treated as commuting coordinates, while analogs of our
$q_i$'s were considered as conjugate canonical (i.e., commuting)
momenta, and the integrals of motion similar to ours $\bar{\Pi}_i$,
and to be analogous to the NR-anyon integrals ${\cal V}_i$, were
ignored there (hence, see footnote 3).

We constructed and analyzed the \emph{nonrelativistic}
\emph{Ex}-model associated with the systems of the EM-particle and
NR-anyon proceeding from the similarity of their dynamics. A similar
relation between dynamics of the relativistic particle in a
homogeneous background electromagnetic field and of the relativistic
model of the particle with torsion \cite{Pol,Ptor}, underlying the
models for relativistic anyons \cite{RA,JN1,CP}, was observed in
\cite{Peq}. Therefore, the relation between these two relativistic
systems can be investigated in the analogous way. Then, one can
expect that the corresponding ``unifying'' extended exotic
relativistic system should produce the nonrelativistic {\it
Ex}-model under application to it of the Jackiw-Nair limit
\cite{JN,HP2}.

The ideas of the present research, being generalized for the field
case, could  be applied to the nonrelativistic Chern-Simons gauge
theory \cite{JPi} playing a prominent role in the description of
anyons and fractional quantum Hall effect (see \cite{Ezw} and
references therein), and to the doubly special relativity
\cite{DSR}.

\vskip 0.4cm\noindent {\bf Acknowledgements}. MP is indebted to
the University of Valladolid for hospitality extended to him, and
to J. L. Cortes, A. Ferrandez and J. Negro for stimulating
discussions. The authors thank R. Jackiw for communication. This
work has been partially supported by the FONDECYT-Chile (projects
1010073, 7010073, 1050001 and 7050046), by the DGI of the
Ministerio de Educaci\'on y Ciencia of Spain (project
BMF2002-02000), by the FEDER Programme of the European Community,
and by the Junta de Castilla y Le\'on (project VA013C05).



\end{document}